# Similarities and Distinctions in Cosmic-Ray Modulation during Different Phases of Solar and Magnetic Activity Cycles

O.P.M. Aslam · Badruddin

Department of Physics, Aligarh Muslim University, Aligarh-202002, India.
e-mail: badr.physamu@gmail.com

**Abstract** We study the solar-activity and solar-polarity dependence of galactic cosmic-ray intensity (CRI) on the solar and heliospheric parameters playing a significant role in solar modulation. We utilize the data for cosmic-ray intensity as measured by neutron monitors, solar activity as measured by sunspot number (SSN), interplanetary plasma/field parameters, solar-wind velocity [$V$] and magnetic field [$B$], as well as the tilt of the heliospheric current sheet [ ] and analyse these data for Solar Cycles 20 - 24 (1965 – 2011). We divide individual Solar Cycles into four phases, *i.e.* low, high, increasing, and decreasing solar activity. We perform regression analysis to calculate and compare the CRI-response to changes in different solar/interplanetary parameters during (i) different phases of solar activity and (ii) similar activity phases but different polarity states. We find that the CRI-response is different during negative ($A<0$) as compared to positive ($A>0$) polarity states not only with SSN and    but also with $B$ and $V$. The relative CRI-response to changes in various parameters, in negative ($A<0$) as compared to positive ($A>0$) state, is solar-activity dependent; it is  2 to 3 times higher in low solar activity,  1.5 to 2 times higher in moderate (increasing/decreasing) activity, and it is nearly equal in high solar-activity conditions. Although our results can be ascribed to preferential entry of charged particles via the equatorial/polar regions of the heliosphere as predicted by drift models, these results also suggest that we should look for, any polarity-dependent response of solar wind and transport parameters in modulating CRI in the heliosphere.

## 1. Introduction

Cosmic rays are modulated as they traverse into the heliosphere. Measurements have shown that cosmic-ray intensity varies on different time scales. On a longer time scale, the cosmic-ray intensity variations in anti-phase with solar activity having  11-year cyclicity is well known and studied extensively *e.g.,* see (Venkatesan and Badruddin, 1990; Storini *et al.,* 1995; Usoskin *et al.,* 1998; Mavromichalaki, Belehaki, and Rafios, 1998; Kane, 2003; Sabbah and Rybansky, 2006; Ma, Han, and Yin*,* 2009; Chaudhary, Dwivedi, and Ray 2011; Kudela, 2012). In addition, the  22-year modulation cycle in cosmic-ray flux due to solar magnetic field polarity reversals is observed and modeled (*e.g.,* Jokipii and Thomas, 1981; Potgieter and Moraal, 1985; Webber and Lockwood, 1988; Potgieter, 1995; Cliver and Ling, 2001; Laurenza *et al.,* 2012). However, in order to understand these phenomena, it is important to understand how cosmic-ray transport and propagation vary with the solar activity and solar magnetic polarity (Heber, 2013).

Cosmic-ray transport in the heliosphere is described by the Parker transport equation (Parker, 1965). If $f(r, P, t)$ is the cosmic-ray distribution with respect to particle rigidity [$P$], then the cosmic ray variation with time [$t$] and position [$r$] is given by

$$\frac{\partial f}{\partial t} = -\mathbf{V} \cdot \nabla f - <v_D> \cdot \nabla f + \nabla \cdot (k_s \cdot \nabla f) + \frac{1}{3} \nabla \cdot \mathbf{V}) \frac{\partial f}{\partial \ln P} \qquad (1)$$

The first term on the right-hand side represents an outward convection caused by the solar-wind velocity [*V*]. The second term represents the gradient and curvature drifts in the global heliospheric magnetic field. The drift velocity for weak scattering is given by

$$< v_\mathrm{D} > = f_s \frac{vpc}{3q} \nabla \times \frac{B}{B^2} \qquad (2)$$

The third term represents the diffusion caused by turbulent irregularities in the background heliospheric magnetic field [*B*]. The last term describes the adiabatic energy change depending on the sign of the divergence of solar-wind velocity [*V*]. Thus we see that in each of the four terms, directly or indirectly, either solar-wind velocity [*V*] or heliospheric magnetic field [*B*] is involved (see, *e.g.*, reviews by, Heber, 2013; Kota, 2013; Strauss, Potgieter, and Ferreira, 2012 and references therein).

Although sunspots have been used as a convenient index of solar activity, anti-correlated with cosmic-ray intensity, it has been recognized that they are not intrinsically related to the problem of solar modulation of cosmic rays. However, the diffusion coefficient changes during 11-year sunspot cycle, being smaller during periods of high solar activity than during periods of low solar activity.

Change in the tilt [ ] of the heliosphere current sheet (HCS) is an important index used for the study of drift effects. However, it is not just the heliospheric current-sheet tilt angle that matters but global gradient and curvature drift effects dependent on the polarity of the global solar magnetic field. The effects of drifts on cosmic rays are such that positively charged cosmic ray particles drift primarily from the polar regions towards lower latitudes and outward along the HCS in the *A*>0 polarity state. In the *A*<0 states, the positive particle-drift directions essentially reverse, and they drift inward along the HCS and then up to higher latitudes. For negatively charged particles, the drift directions are opposite to those of positively charged particles in any given polarity state of the heliosphere. There are indications that cosmic-ray intensity decreases more rapidly as sunspot number (SSN) increases during the increasing phase of solar cycles when solar polarity is negative (*A*<0) than when *A* is positive (*A*>0) (Van Allen 2002; Singh, Singh, and Badruddin, 2008). Similar differences in cosmic-ray changes in *A*<0 and *A*>0 solar-polarity states have been observed with tilt angle [ ] changes also (Smith and Thomas, 1986; Webber and Lockwood, 1988; Smith, 1990; Badruddin, Singh, and Singh, 2007). Although the both parameters SSN and    indicate their different effectiveness in modulating cosmic-ray intensity in different polarity conditions of the heliosphere, it will be interesting to explore whether the cosmic-ray response is different with the parameters *V* and *B* also, in different polarity states of the heliosphere (*A*<0 and *A*>0). Although there is some indication of such effects (with *V*, *B*) during solar minimum (Richardson, Cane, and Wibberenz, 1999; Badruddin, 2011; Cliver, Richardson, and Ling, 2013; Aslam and Badruddin, 2012), it will be of special significance if a such study is done during different solar-activity conditions (low, increasing, decreasing, and high) and different polarity states (*A*<0 and *A*>0), in view of the current paradigm of cosmic-ray transport (Jokipii and Wibberenz, 1998; Kota, 2013), *i.e.* particle drifts play an important role during low to moderate solar activity, while solar maxima are dominated by large-scale diffusion barriers, called global merged interaction regions (GMIRs), sweeping out the cosmic rays (Burlaga *et al.,* 1985; McDonald, Lal, and McGuire, 1993). It remains an open question whether drifts are fully turned off during solar maximum, or they are simply there but masked by other, more aggressive processes (Kota, 2013).

We have recently studied the CRI modulation during the deep minimum of Solar Cycle 23 (Badruddin, 2011) and declining including minimum phase of the four Solar Cycles 20 - 23 (Aslam and Badruddin, 2012). This period (declining and minimum solar activity) corresponds to the recovery phase of the 11-year cosmic-ray modulation cycle, after intensity depression (modulation) during increasing and maximum phases of the corresponding solar-activity cycle. In this article, we study the cosmic-ray flux variability with SSN and    as well as with *V* and *B* and two derivatives of *V* and *B*, *i.e.* interplanetary electric field ($E = BV/1000$ mV m$^{-1}$) and $BV^2$ [mV s$^{-1}$], considered important for solar modulation of cosmic rays (Sabbah 2000; Ahluwalia 2005; Sabbah and Kudela, 2011). This study has been done not only during increasing including maximum phases of different Solar Cycles but also during four activity phases (low, increasing, decreasing, and high) of Solar Cycles 20 – 23, and the increasing phase of Cycle 24 (up to 2011).

## 2. Results and Discussion

In Figure 1, we have plotted solar [SSN] and interplanetary parameters [$V$, $B$], standard deviation of field vector [$B$ ($\sigma_B$)], their derivatives [$BV/1000$, $BV^2$] (omniweb.gsfc.nasa.gov), tilt angle [ ; wso.stanford.edu] and cosmic ray intensity [CRI: cosmicrays.oulu.fi] for Solar Cycles 20, 21, 22, 23, and 24 (up to 2011). This figure is shown to highlight the nature of changing activity in different solar cycles and the nature of simultaneous variations in various parameters considered for the analysis presented in this article.

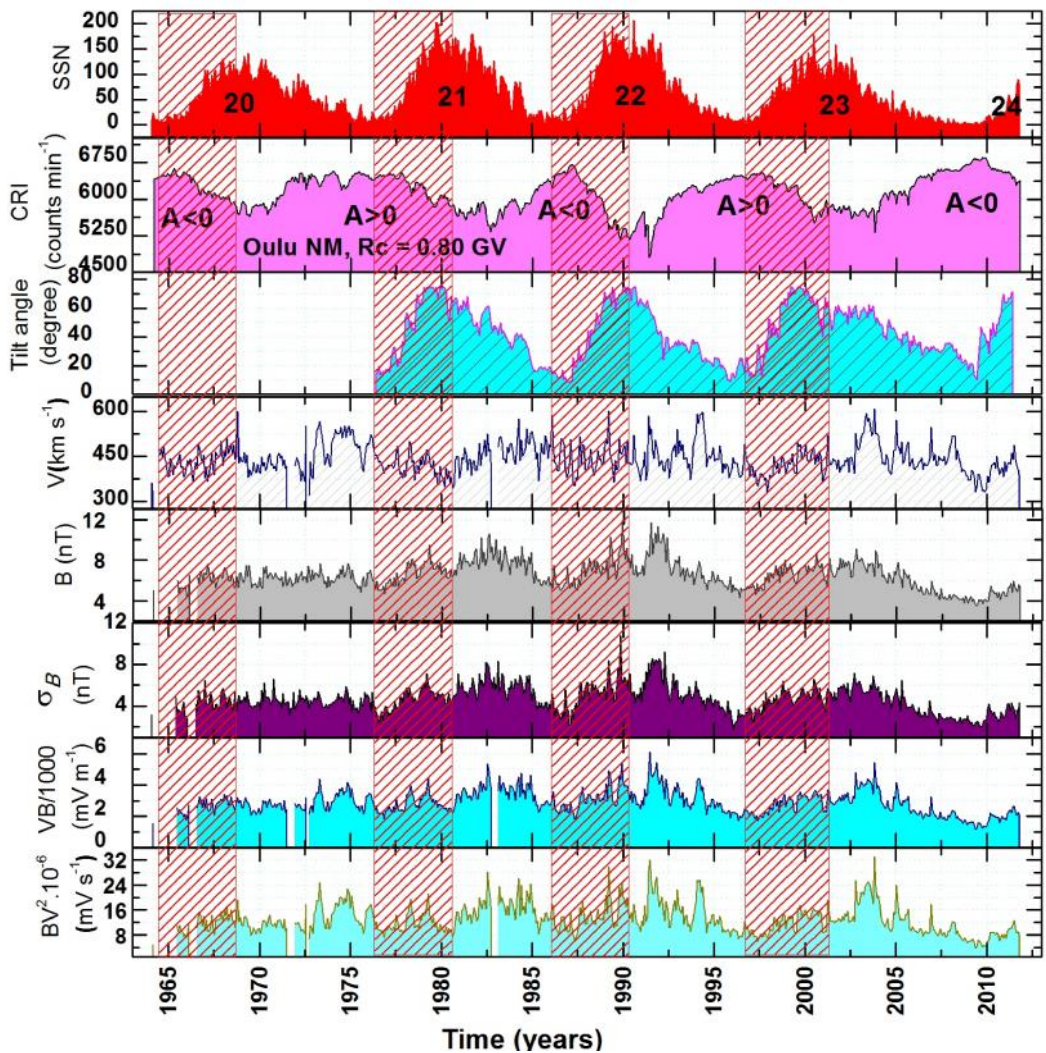

**Figure 1** Solar-rotation-averaged solar [SSN] and interplanetary [$V$, $B$] parameters, standard deviation of the vector field [$\sigma_B$], their derivatives [$BV/1000$, $BV^2$], tilt angle [ ] and cosmic-ray intensity for Solar Cycles 20, 21, 22, 23, and 24 (up to 2011), increasing and including maximum phase of solar cycles are hatched in red.

For the analysis we considered, at first, the increasing including maximum phase of Solar Cycles 20, 21, 22, and 23 (see Figure 1; shaded portions); this is mainly the period when the CRI depression (modulation) takes place before it starts recovering to its pre-decrease level to complete the cycle.

Differences in CRI profiles during alternate polarity cycles are well known; these differences are considered to be the consequence of drift effects (Jokipii and Thomas, 1981; Venkatesan and Badruddin, 1990; Potgieter, 1995; Jokipii and Wibberenz, 1998; Kota, 2013 and references therein). As shown in Figure 1, there are large temporal variations in 27-day averages of different parameters. Thus, information that is more useful may be obtained if we compare the variations in CRI during similar phases of different Solar Cycles with variations in relevant solar and interplanetary parameters.

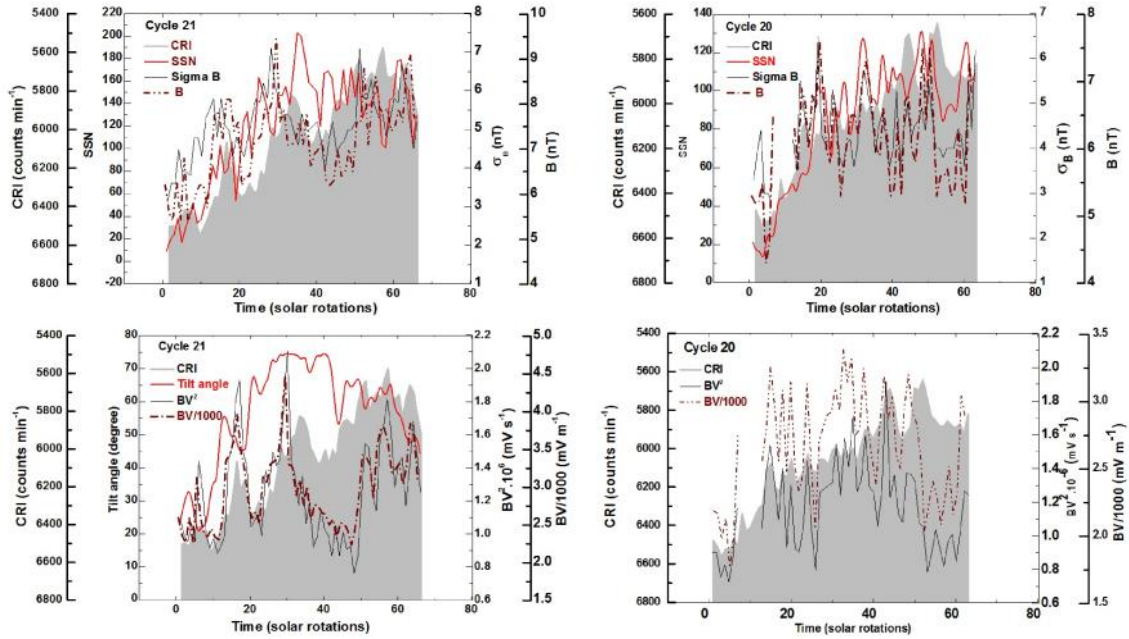

**Figure 2a** Comparison of 27-day averages of various solar and interplanetary parameters during increasing, including maximum, phase of Solar Cycles 20 (right panel) and 21 (left panel). Note CRI (gray shaded) scale is inverted in these figures.

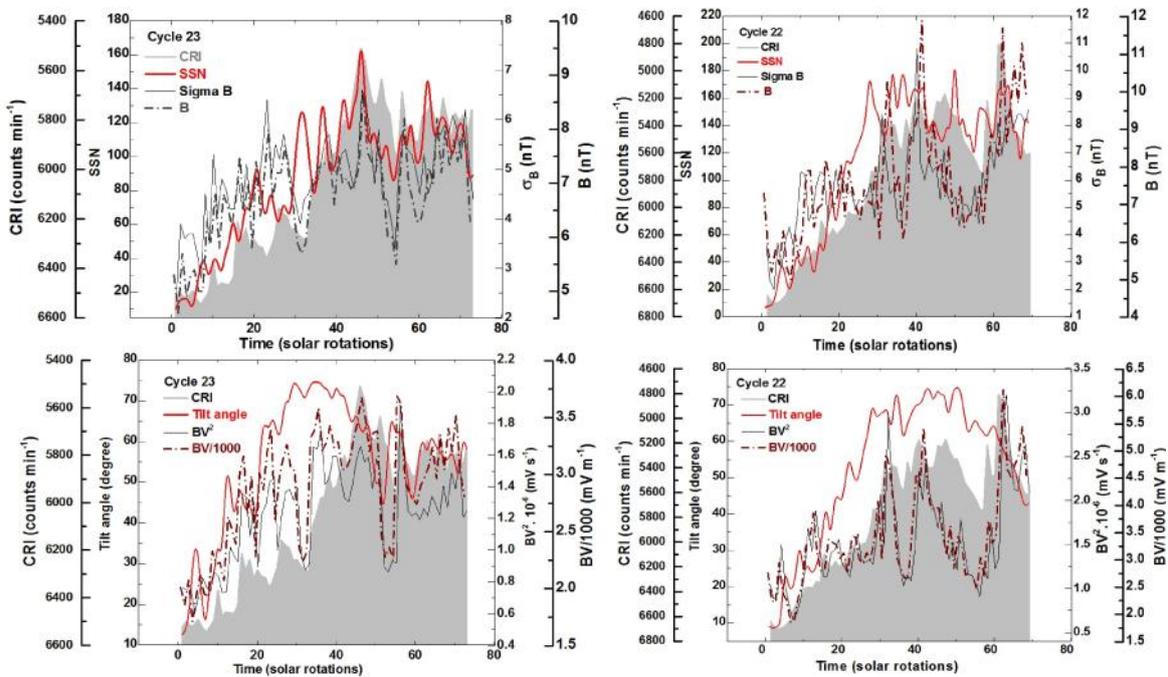

**Figure 2b** Temporal variations of CRI and various solar/interplanetary parameters during increasing including maximum phase of Solar Cycles 22 (right panel) and 23 (left panel). Note CRI (gray hatching) scale is inverted in these figures.

In Figure 2a and 2b) we compare, during the increasing and including maximum phase of Solar Cycles 20, 21, 22, and 23, 27-day averages of various solar and interplanetary parameters [SSN, $B$, $\sigma_B$, , $E$, and $BV^2$) with variations in CRI. Note that the cosmic-ray intensity scale is inverted for better comparison. Although the trends in changes in CRI appear to match with many of the parameters considered, step changes in CRI do not appear to track well with many of these parameters, and also, not consistently similar in all the four Solar Cycles. Although these figures provide a general trend for comparison of changes in CRI and various solar/interplanetary parameters during the increasing and including maximum phase of different Solar Cycles 20 - 23, they do not provide us with a quantitative information about relatively poorly or better related parameters.

**Table 1** Rate of change of cosmic-ray intensity with various parameters [$\Delta I/\Delta P$] and correlation coefficient [$R$] between cosmic ray intensity and different parameters after during increasing, including maximum, phases of Solar Cycles 20, 21, 22, and 23. The time lag (solar rotations) between CRI and various parameters are also given.

| Parameters | Solar Cycle 20 ($A<0$) | | | Solar Cycle 21 ($A>0$) | | | Solar Cycle 22 ($A<0$) | | | Solar Cycle 23 ($A>0$) | | |
|---|---|---|---|---|---|---|---|---|---|---|---|---|
| | Lag | $\Delta I/\Delta P$ | $R$ | Lag | $\Delta I/\Delta P$ | $R$ | Lag | $\Delta I/\Delta P$ | $R$ | Lag | $\Delta I/\Delta P$ | $R$ |
| SSN | 3 | -5.76 ± 0.41 | -0.87 | 14 | -4.1 ± 0.25 | -0.90 | 5 | -7.59 ± 0.45 | -0.90 | 5 | -6.15 ± 0.41 | -0.89 |
| Tilt angle [degree] | | | | 15 | 10.3 ± 0.79 | 0.86 | 6 | 20.1 ± 1.06 | 0.92 | 15 | 12.4 ± 0.71 | 0.90 |
| $V$ [km s$^{-1}$] | 0 | -1.72 ± 0.74 | -0.29 | 0 | -0.29 ± 0.98 | -0.04 | 0 | -3.61 ± 1.1 | -0.37 | 0 | -4.65 ± 0.80 | -0.57 |
| $B$ [nT] | 0 | -120.4 ± 37.7 | -0.41 | 0 | -188 ± 28.3 | -0.64 | 0 | -204.6 ± 30.5 | -0.66 | 0 | -250.4 ± 21.1 | -0.81 |
| $\sigma_B$ [nT] | 0 | -101.2 ± 37.7 | -0.34 | 0 | -202.4 ± 31.1 | -0.63 | 0 | -201 ± 28.98 | -0.65 | 0 | -231.5 ± 20.3 | -0.80 |
| $BV/1000$ [mV m$^{-1}$] | 0 | -258.8 ± 69.2 | -0.46 | 0 | -249.0 ± 62.6 | -0.45 | 0 | -361.2 ± 51.4 | -0.65 | 0 | -407.0 ± 50.5 | -0.69 |
| $BV^2$ [$10^{-6}$ mV s$^{-1}$] | 0 | -311.3 ± 84.4 | -0.46 | 0 | -271.3 ± 11.6 | -0.30 | 0 | -397.3 ± 82.4 | -0.56 | 1 | -592.4 ± 84.2 | -0.64 |

A careful examination of Figure 2a and 2b shows that there appears to be a slight shift in the occurrence of steps in CRI changes and some of the solar/interplanetary parameters. This may be due to some time lag between the changes in CRI and these solar/interplanetary parameters. Thus, we have determined the time lag between the solar rotation averaged CRI and various solar/interplanetary parameters during increasing including maximum phase of Solar Cycles 20, 21, 22, and 23 (see Table 1). We found that, at the time scale of solar rotations, there is no (zero) time lag between CRI and interplanetary plasma/field parameters [*e.g.*, *V, B*]. However, the lag between CRI and SSN/ varies from three to 15 solar rotations during increasing including maximum phase of different Solar Cycles. After introducing the time lag, wherever applicable as given in Table 1, we calculate the rate of change of CRI with changes in different parameters ($\Delta I/\Delta P$). The calculated values of $\Delta I/\Delta P$, and the correlation coefficients ($R$) between CRI and various parameters are given in Table 1.

We observe from this table that, although there is some indication that the CRI decreases at a faster rate with increase in certain parameters, (*e.g.* HCS tilt) during $A<0$ as compared to $A>0$, however, such a difference is not consistently seen with all the parameters when $\Delta I/\Delta P$ is calculated during the increasing and including maximum phase of different Solar Cycles (20, 21, 22, and 23). Such a difference may be expected if drifts were a dominant effect in solar modulation during increasing including maximum phase of different solar cycles. However, it is also likely that drifts are fully turned off during solar maximum or they are there but masked by other more aggressive processes (Kota, 2013). If such is the case, then excluding the solar maximum and performing analysis by considering only the increasing phase of different solar cycles is likely to be more informative for gaining insight about the modulation process.

Table 2 Rate of change of cosmic-ray intensity (Oulu NM) with various parameters [$\Delta I/\Delta P$] and correlation coefficient [$R$] between different parameters during increasing phases of Solar Cycles 21, 22, 23, and 24.

| Parameters | Solar Cycle 21 ($A>0$) | | Solar Cycle 22 ($A<0$) | | Solar Cycle 23 ($A>0$) | | Solar Cycle 24 ($A<0$) | |
|---|---|---|---|---|---|---|---|---|
| | $\Delta I/\Delta P$ | $R$ | $\Delta I/\Delta P$ | $R$ | $\Delta I/\Delta P$ | $R$ | $\Delta I/\Delta P$ | $R$ |
| SSN | -3.2 ± 0.37 | -0.85 | -6.45 ± 0.50 | -0.92 | -3.83 ± 0.44 | -0.85 | -9.6 ± 1.22 | -0.82 |
| Tilt angle [degree] | -7.92 ± 0.74 | -0.89 | -17.1 ± 1.18 | -0.94 | -5.9 ± 0.55 | -0.89 | -9.69 ± 0.77 | -0.92 |
| $V$ [km s$^{-1}$] | -2.6 ± 0.85 | -0.49 | -3.39 ± 0.96 | -0.55 | -2.75 ± 0.52 | -0.70 | -3.67 ± 0.62 | -0.74 |
| $B$ [nT] | -140.1 ± 16.9 | -0.84 | -221.2 ± 43.5 | -0.69 | -107.9 ± 15.1 | -0.80 | -253.7 ± 36.8 | -0.79 |
| $\sigma_B$ [nT] | -154.7 ± 15.2 | -0.88 | -189.2 ± 29.4 | -0.77 | -111.3 ± 15.9 | -0.79 | -210.4 ± 36.2 | -0.73 |
| $BV/1000$ [mV m$^{-1}$] | 214.4 ± 31.1 | 0.82 | 392.6 ± 64.0 | 0.75 | 210.7 ± 26.6 | 0.83 | 426.1 ± 51.4 | 0.84 |
| $BV^2$ [10$^{-6}$ mV s$^{-1}$] | -403.2 ± 63.4 | -0.76 | -531.6 ± 96.9 | -0.71 | -369.8 ± 51.3 | -0.80 | -733.2 ± 94.2 | -0.82 |

Cliver, Richardson, and Ling (2013) remarked that the effects of gradient and curvature drifts are most notable at the onset of modulation cycles in $A>0$ epochs when CRI responds weakly to increase in $B$ and . The weak response of the CRI to changes in $B$ during the rise of odd numbered cycles ($A>0$ epochs) is attributed to drift-induced preference for positively charged particles to approach the inner heliosphere from the poles at these times (Jokipii and Thomas, 1981) and the relative confinement of coronal mass ejections to low latitudes at the onset of the Solar (modulation) Cycle (Gopalswamy *et al.*, 2010). On the other hand, their observations (see Cliver, Richardson, and Ling, 2013) challenge the pre-eminence of drifts during the recent solar minimum following Cycle 23, which is not in line with the current paradigm for the modulation of CRI, *i.e.* diffusion is the dominant process during the solar maximum while drift dominates at minima.

Table 3 Rate of change of cosmic-ray intensity (Newark NM) with various parameters [$\Delta I/\Delta P$] and correlation coefficient [$R$] between different parameters during increasing phases of Solar Cycles 21, 22, 23, and 24.

| Parameters | Solar cycle 21 ($A>0$) | | Solar cycle 22 ($A<0$) | | Solar cycle 23 ($A>0$) | | Solar cycle 24 ($A<0$) | |
|---|---|---|---|---|---|---|---|---|
| | $\Delta I/\Delta P$ | $R$ | $\Delta I/\Delta P$ | $R$ | $\Delta I/\Delta P$ | $R$ | $\Delta I/\Delta P$ | $R$ |
| SSN | -1.64 ± 0.20 | -0.83 | -3.04 ± 0.24 | -0.92 | -2.01 ± 0.24 | -0.84 | -5.34 ± 0.59 | -0.86 |
| Tilt angle [degree] | -4.16 ± 0.39 | -0.89 | -8.12 ± 0.55 | -0.94 | -3.16 ± 0.27 | -0.91 | -5.5 ± 0.39 | -0.94 |
| $V$ [km s$^{-1}$] | -1.39 ± 0.44 | -0.50 | -1.58 ± 0.46 | -0.54 | -1.49 ± 0.26 | -0.72 | -2.21 ± 0.28 | -0.82 |
| $B$ [nT] | -71.0 ± 9.6 | -0.81 | -108.3 ± 20.2 | -0.71 | -60.8 ± 7.8 | -0.82 | -129.7 ± 18.6 | -0.80 |
| $\sigma_B$ [nT] | -76.9 ± 9.4 | -0.84 | -91.7 ± 13.5 | -0.78 | -59.1 ± 8.3 | -0.80 | -117.5 ± 18.2 | -0.77 |
| $BV/1000$ [mV m$^{-1}$] | -127 ± 16.6 | -0.82 | -185.3 ± 30.4 | -0.75 | -113.6 ± 13.3 | -0.85 | -225.2 ± 26.1 | -0.85 |
| $BV^2$ [10$^{-6}$ mV s$^{-1}$] | -212.3 ± 32.4 | -0.77 | -254.3 ± 46.4 | -0.71 | -213.2 ± 26.8 | -0.82 | -412.3 ± 45.7 | -0.86 |

Thus, we consider first only the increasing phase of Solar Cycle 24 (30 rotations after the minimum of Solar Cycle 23) and calculate the rate of change in Oulu CRI (cutoff rigidity Rc = 0.80 GV,

Latitude = 65.05° N and Longitude = 25.47° E) (Table 2) with different parameters [$\Delta I/\Delta P$] and compared these results with a similar period of the three previous Solar Cycles 21, 22, and 23. To check the consistency of these results, a similar analysis for the same periods is done for another neutron monitor (Newark; neutronm.bartol.udel.edu) with cutoff rigidity Rc = 2.09 GV, Latitude = 39.7° N and Longitude = 75.7° W; these results are also tabulated (see Table 3). A careful examination of the values of $\Delta I/\Delta P$ at both of the neutron monitors (Tables 2 and 3) shows that the CRI decreases at a faster rate with an increase in almost all of the parameters considered in this analysis [SSN, $\alpha$, V, B, $\sigma_B$, E, and $BV^2$] in the A<0 polarity epoch. In fact the decrease in CRI with increase in various parameters is 1.5 to 2 times faster during increasing phase, in A<0 polarity state (Cycle 22 and 24) as compared to A>0 polarity state (Cycle 21 and 23) (see Table 4).

Table 4 Comparison of rate of change of CRI with different parameters [$\Delta I/\Delta P$] during A<0 and A>0 polarity epochs during the increasing phase of Solar Cycles 21 - 24.

| Parameters | <$\Delta I/\Delta P$> | | <$\Delta I/\Delta P$> | | Ratio | |
|---|---|---|---|---|---|---|
| | Oulu NM (A<0) | Oulu NM (A>0) | Newark NM (A<0) | Newark NM (A>0) | Oulu NM (A<0):(A>0) | Newark NM (A<0):(A>0) |
| SSN | -8.03 ± 0.86 | -3.52 ± 0.40 | -4.19 ± 0.42 | -1.83 ± 0.22 | ≈ 2 : 1 | ≈ 2 : 1 |
| Tilt angle [degree] | -13.4 ± 0.98 | -6.92 ± 0.65 | -6.81 ± 0.47 | -3.66 ± 0.33 | ≈ 2 : 1 | ≈ 2 : 1 |
| V [km s$^{-1}$] | -3.53 ± 0.79 | -2.68 ± 0.69 | -1.9 ± 0.37 | -1.44 ± 0.7 | ≈ 3 : 2 | ≈ 3 : 2 |
| B [nT] | -237.4 ± 40.2 | -124.0 ± 16.0 | -119.0 ± 19.4 | -65.9 ± 8.7 | ≈ 2 : 1 | ≈ 2 : 1 |
| $\sigma_B$ [nT] | -199.8 ± 32.9 | -133.0 ± 15.6 | -104.6 ± 15.9 | -68.0 ± 8.85 | ≈ 3 : 2 | ≈ 3 : 2 |
| BV/1000 [mV m$^{-1}$] | -409.3 ± 57.7 | -226.55 ± 28.85 | -205.3 ± 28.3 | -120.3 ± 14.95 | ≈ 2 : 1 | ≈ 2 : 1 |
| $BV^2$ [10$^6$.mVs$^{-1}$] | -632.4 ± 95.6 | -386.5 ± 57.4 | -333.3 ± 46.1 | -212.8 ± 29.6 | ≈ 3 : 2 | ≈ 3 : 2 |

Table 5a. Rate of change of cosmic-ray intensity with various parameters [$\Delta I/\Delta P$] and correlation coefficient [R] between different parameters during different low activity (minimum) periods of Solar Cycles.

| Parameters | Oulu NM | | | | Newark NM | | | |
|---|---|---|---|---|---|---|---|---|
| | 2007 - 09 (Solar Cycle 23) (A<0) | | 1995 - 97 (Solar Cycle 22) (A>0) | | 2007 - 09 (Solar Cycle 23) (A<0) | | 1995 - 97 (Solar Cycle 22) (A>0) | |
| | $\Delta I/\Delta P$ | R | $\Delta I/\Delta P$ | R | $\Delta I/\Delta P$ | R | $\Delta I/\Delta P$ | R |
| SSN | -10.3 ± 1.9 | -0.66 | -3.9 ± 0.72 | -0.66 | -4.45 ± 0.73 | -0.70 | -1.71 ± 0.31 | -0.73 |
| Tilt angle [degree] | -17.2 ± 1.3 | -0.90 | -5.9 ± 0.66 | -0.82 | -6.0 ± 0.52 | -0.88 | -1.87 ± 0.56 | -0.57 |
| V [km s$^{-1}$] | -2.03 ± 0.24 | -0.80 | -1.0 ± 0.21 | -0.61 | -0.76 ± 0.08 | -0.85 | -0.40 ± 0.09 | -0.56 |
| B [nT] | -174.7 ± 33.5 | -0.64 | -51.2 ± 8.5 | -0.69 | -65.33 ± 12.42 | -0.64 | -24.16 ± 6.05 | -0.54 |
| $\sigma_B$ [nT] | -186.6 ± 33.6 | -0.66 | -42.9 ± 6.0 | -0.75 | -64.42 ± 12.3 | -0.64 | -19.23 ± 3.4 | -0.67 |
| BV/1000 [mV m$^{-1}$] | -322.5 ± 38.0 | -0.81 | -63.8 ± 8.95 | -0.75 | -104.2 ± 11.16 | -0.83 | -43.44 ± 10.63 | -0.66 |
| $BV^2$ [10$^6$.mVs$^{-1}$] | -458.3 ± 57.4 | -0.79 | -83.2 ± 12.4 | -0.73 | -152.2 ± 16.3 | -0.84 | -69.7 ± 17.4 | -0.66 |

**Figure 3a** Change in CRI with SSN during different (low, increasing, decreasing, and high) solar-activity periods in two polarity states ($A<0$ and $A>0$). Best-fit linear curves along with correlation coefficients are also given in each case.

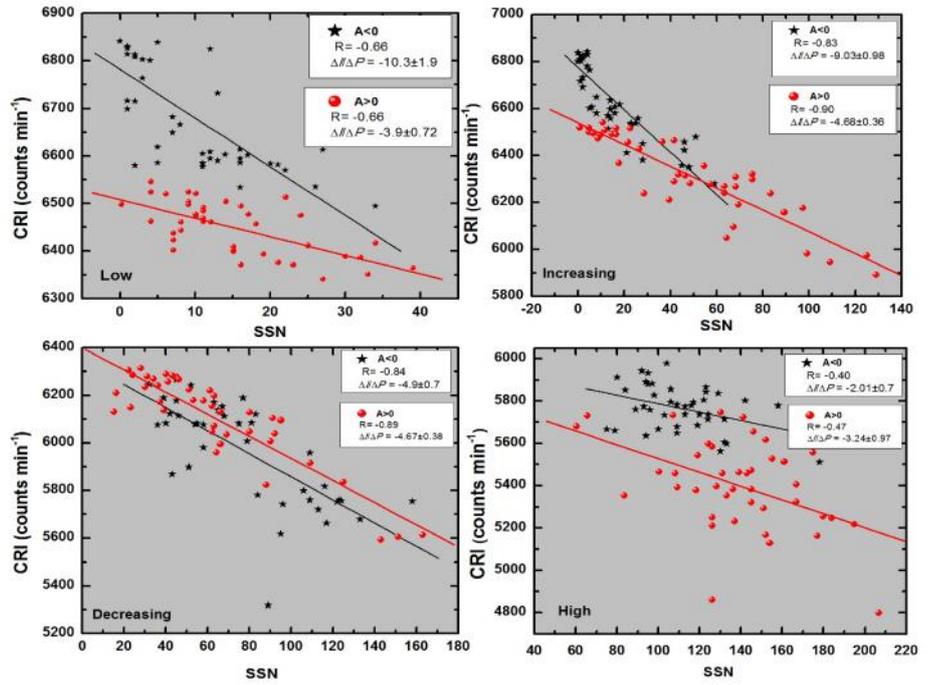

There is some indication that the response of the cosmic rays to solar-wind speed changes during solar minima of different polarity ($A<0$ and $A>0$) are different (Badruddin, Yadav, and Yadav, 1985; Richardson, Cane, and Wibberenz, 1999; Singh and Badruddin, 2007; Gupta and Badruddin, 2009; Modzelowska and Alania, 2011). For low $B$ values (< 6 nT), in $A>0$ epochs, CRI is reported to decrease more slowly as $B$ increases than in the case for $A<0$ epochs (*e.g.* Wibberenz, Richardson, and Cane, 2002; Cliver, Richardson, and Ling, 2013). There were also suggestions for the polarity dependence of the transport parameters such as parallel mean free path [ ], with its value being substantially larger during solar minimum periods with negative polarity ($A<0$) than in those with positive polarity (Chen and Bieber, 1993). However, the implications of these results are not in agreement with those of drift model calculations (*e.g.* see Kota and Jokipii, 1991; Richardson, Cane, and Wibberenz, 1999).

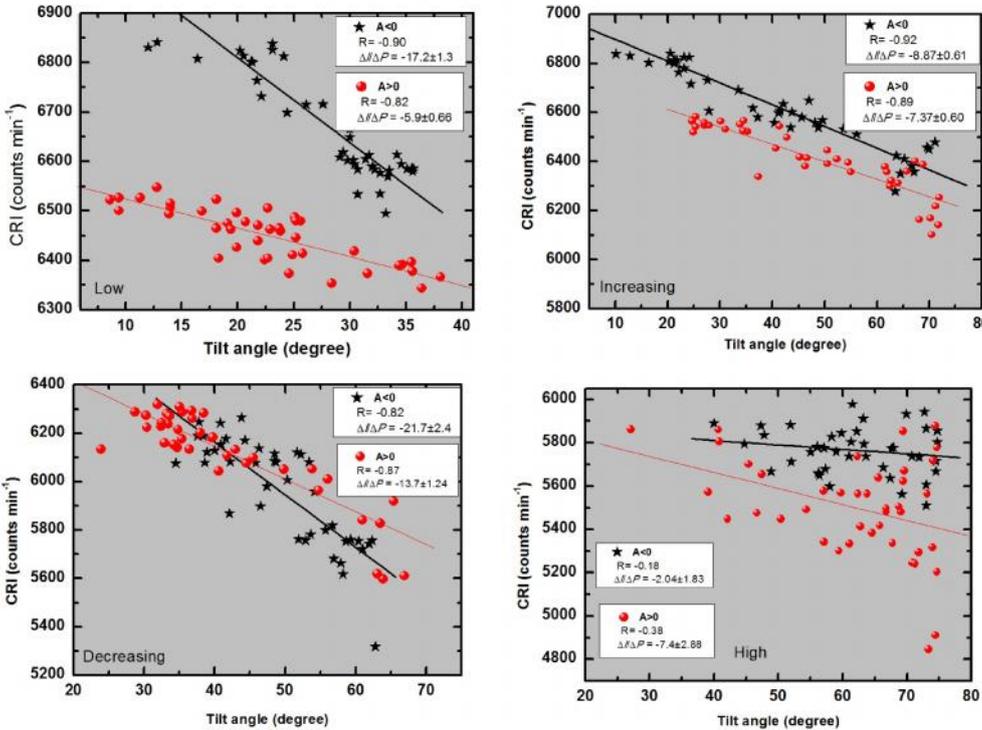

**Figure 3b** Change in CRI with respect to heliospheric current sheet tilt [ ] during different (low, increasing, decreasing, and high) solar activity periods and polarity states of the heliosphere.

The reduced responsiveness of cosmic rays to sunspots or tilt angle increases during $A>0$ epochs in the increasing phase of Solar Cycles as compared to the $A<0$ epoch are considered to be supportive of drift models of CRI modulation (Smith and Thomas, 1986; Webber and Lockwood, 1988; Smith, 1990; Van Allen, 2000; Cliver and Ling, 2001; Badruddin, Singh, and Singh, 2007; Singh, Singh, and Badruddin, 2008). It needs to be clarified whether similar differences in CRI response to $V$ and $B$ changes in $A<0$ and $A>0$ suggest for any epoch-dependent convection/diffusion effects (transport parameters) in CRI modulation or whether the observed differences in the effectiveness of *V* and *B* too are actually a consequence of different access routes for charged particles in *A*<0 and *A*>0 polarity conditions.

**Table 5b** Rate of change of cosmic-ray intensity with various parameters [$\Delta I/\Delta P$] and correlation coefficient [*R*] between different parameters during different increasing activity periods of Solar Cycles.

| Parameters | Oulu NM | | | | Newark NM | | | |
|---|---|---|---|---|---|---|---|---|
| | 2009 – 11 (Solar Cycle 24) (*A*<0) | | 1997 – 99 (Solar Cycle 23) (*A*>0) | | 2009 – 11 (Solar Cycle 24) (*A*<0) | | 1997 – 99 (Solar Cycle 23) (*A*>0) | |
| | $\Delta I/\Delta P$ | *R* | $\Delta I/\Delta P$ | *R* | $\Delta I/\Delta P$ | *R* | $\Delta I/\Delta P$ | *R* |
| SSN | -9.03 ± 0.98 | -0.83 | -4.68 ± 0.36 | -0.90 | -5.43 ± 0.57 | -0.84 | -2.42 ± 0.19 | -0.90 |
| Tilt [degree] | -8.87 ± 0.61 | -0.92 | -7.37 ± 0.60 | -0.89 | -5.39 ± 0.32 | -0.94 | -3.91 ± 0.29 | -0.90 |
| *V* [km s$^{-1}$] | -3.64 ± 0.53 | -0.74 | -3.52 ± 0.50 | -0.75 | -2.23 ± 0.31 | -0.76 | -1.87 ± 0.25 | -0.76 |
| *B* [nT] | -231.7 ± 27.6 | -0.80 | -149.6 ± 18.5 | -0.79 | -138.29 ± 15.5 | -0.82 | -79.41 ± 9.34 | -0.81 |
| $\sigma_B$ [nT] | -216.6 ± 27.9 | -0.78 | -145.2 ± 16.9 | -0.81 | -127.41 ± 17.06 | -0.77 | -77.2 ± 8.5 | -0.83 |
| *BV*/1000 [mV m$^{-1}$] | -403.3 ± 40.8 | -0.85 | -278.3 ± 31.8 | -0.81 | -254.6 ± 20.76 | -0.89 | -142.95 ± 17.05 | -0.80 |
| $BV^2$ [10$^{-6}$ mV s$^{-1}$] | -713.7 ± 74.1 | -0.84 | -479.6 ± 52.3 | -0.83 | -448.6 ± 39.3 | -0.88 | -252.4 ± 27.4 | -0.82 |

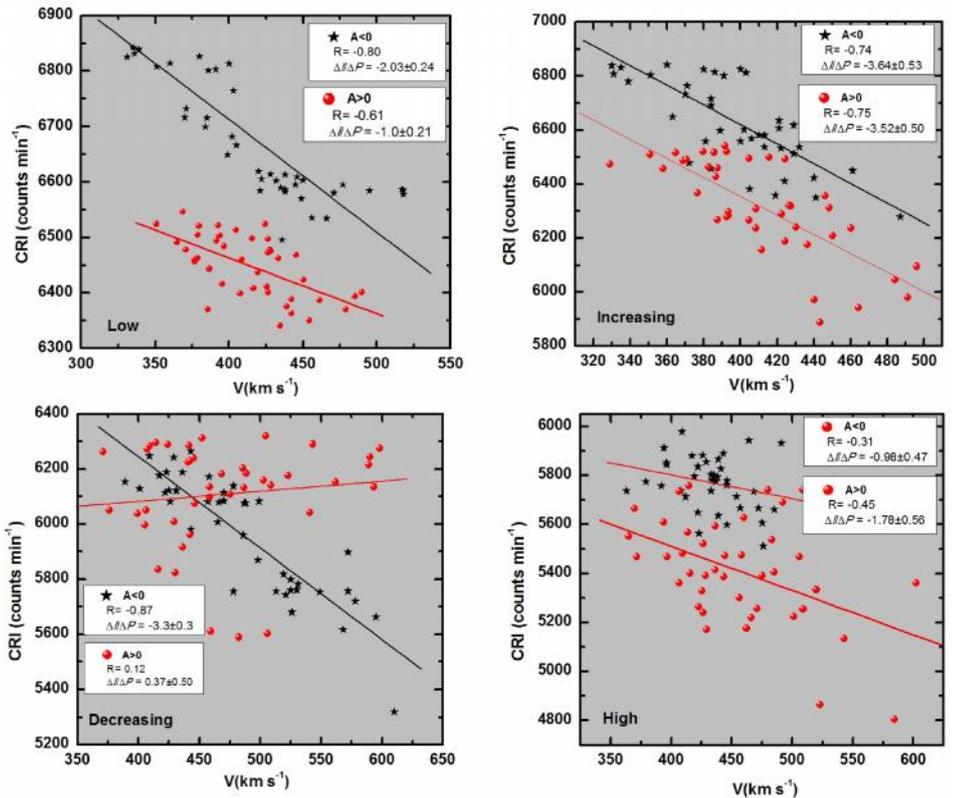

**Figure 3c** Change in CRI with solar wind velocity [*V*] during different (low, increasing, decreasing, and high) solar activity periods in two polarity states (*A*<0 and *A*>0).

It will be interesting to see whether such differences in $\Delta I/\Delta P$ in *A*<0 and *A*>0 polarity epochs are similar in nature and magnitude or different during different levels/phases of solar cycles. From visual inspection of sunspot cycles, we have divided the solar cycles into four parts; increasing, maximum (high), decreasing, and minimum (low) solar-activity periods. The periods considered for this particular analysis are 1989 - 1991 (maximum), 1992 - 1994 (decreasing), 1995 - 1997 (minimum), 1997 - 1999 (increasing), 2000 - 2002 (maximum), 2003 - 2005 (decreasing), 2007 - 2009 (minimum), and 2009 - 2011 (increasing). In order to keep the number of data points equal for regression analysis, 40 solar rotations (≈ three years) data of each of these parts were used. In this way we have considered two similar phases (*e.g.* low activity) in two different polarity epochs (*A*<0 and *A*>0) of equal duration (40 solar rotations ≈ three years). Similarly, we have also considered two similar phases (*e.g.* increasing and decreasing) of solar activity in two different epochs, one in *A*<0 and other in *A*>0, of the same duration (40 solar rotations) from the latest three Cycles 22, 23, and 24. We have also considered two periods of similar (high) solar activity but mixed polarity. During these periods, the data (especially V and B) are of better quality with fewer data gaps. The scatter plots along with best-fit linear curves are plotted for Oulu NM count rate (see Figure 3a-3d).

Table 5c Rate of cosmic-ray intensity change with various parameters [$\Delta I/\Delta P$] and correlation coefficient [R] between different parameters during different decreasing-activity periods of Solar Cycles.

| Parameters | Oulu NM | | | | Newark NM | | | |
|---|---|---|---|---|---|---|---|---|
| | 2003 - 05 (Solar Cycle 23) (A<0) | | 1992 - 94 (Solar Cycle 22) (A>0) | | 2003 - 05 (Solar Cycle 23) (A<0) | | 1992 - 94 (Solar Cycle 22) (A>0) | |
| | $\Delta I/\Delta P$ | R | $\Delta I/\Delta P$ | R | $\Delta I/\Delta P$ | R | $\Delta I/\Delta P$ | R |
| SSN | -4.9 ± 0.7 | -0.84 | -4.67 ± 0.38 | -0.89 | -3.9 ± 0.61 | -0.82 | -2.52 ± 0.20 | -0.90 |
| Tilt [degree] | -21.7 ± 2.4 | -0.82 | -13.7 ± 1.24 | -0.87 | -10.9 ± 1.3 | -0.80 | -7.74 ± 0.78 | -0.88 |
| V [km s$^{-1}$] | -3.3 ± 0.30 | -0.87 | 0.37 ± 0.5 | 0.12 | -1.81 ± 0.15 | -0.88 | 0.08 ± 0.26 | 0.16 |
| B [nT] | -192.3 ± 22.8 | -0.80 | -109.5 ± 9.84 | -0.87 | -102.5 ± 12.63 | -0.79 | -58.0 ± 5.07 | -0.88 |
| $\sigma_B$ [nT] | -176.2 ± 30.7 | -0.68 | -105.9 ± 11.8 | -0.82 | -93.64 ± 16.8 | -0.67 | -57.6 ± 6.09 | -0.83 |
| BV/1000 [mV m$^{-1}$] | -244.0 ± 20.96 | -0.88 | -181.5 ± 26.5 | -0.74 | -132.36 ± 11.3 | -0.88 | -93.08 ± 14.9 | -0.71 |
| BV$^2$ [10$^{-6}$ mV s$^{-1}$] | -341.7 ± 26.2 | -0.90 | -189.7 ± 53.6 | -0.49 | -182.4 ± 14.6 | -0.90 | -90.8 ± 30.2 | -0.44 |

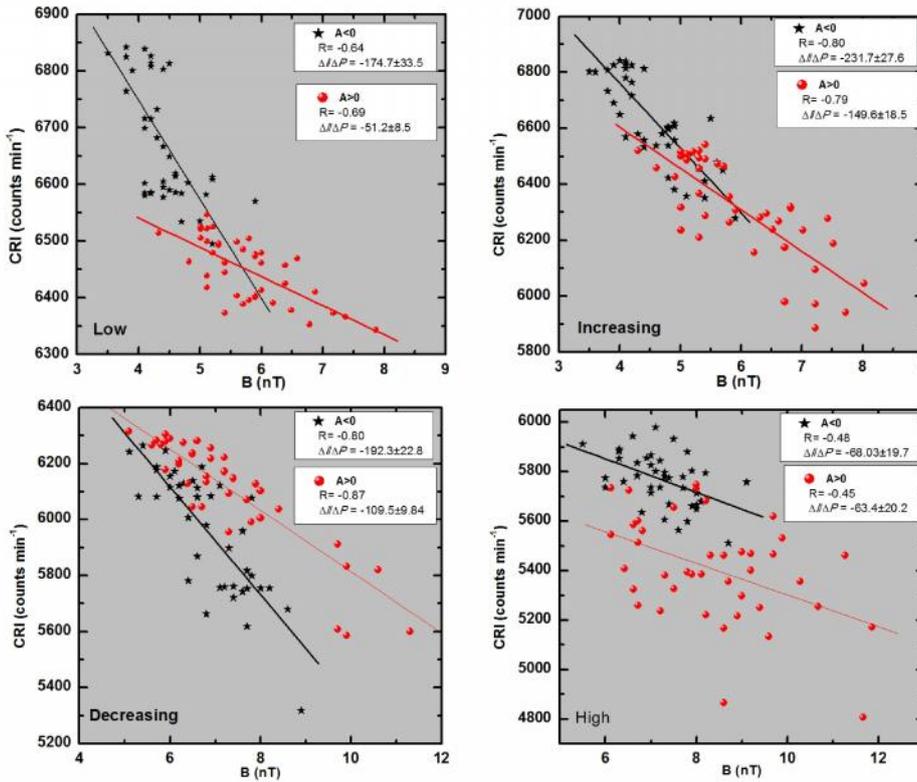

**Figure 3d** Change in CRI with interplanetary magnetic field [B] during different (low, increasing, decreasing, and high) solar activity periods in two polarity states.

To ensure the reliability of the observed results, we have calculated the values of $\Delta I/\Delta P$ during exactly the same periods for another neutron monitor (Newark) located at different latitude and longitude on the earth. These values with correlation coefficients (R) are given in Table 5a (during low activity), Table 5b (during increasing activity), Table 5c (during decreasing activity) and Table 5d (during high solar activity). In this computation, we have taken care that all individual periods are of equal length (40 solar rotations). To quantify how much reduced or enhanced is the CRI-response to changes in various solar/interplanetary parameters, we have calculated the ratio in the values of $\Delta I/\Delta P$ during two polarity states in three solar activity conditions. These values are tabulated in

Table 6. Critical examinations of Figures 3a-3d, Tables 5a-5d and Table 6 lead us to conclude the following.

Table 5d Rate of change of cosmic-ray intensity with various parameters [$\Delta I/\Delta P$] and correlation coefficient [$R$] between different parameters during different high activity (maximum) periods of Solar Cycles.

| Parameters | Oulu NM | | | | Newark NM | | | |
|---|---|---|---|---|---|---|---|---|
| | 2000 - 02 (Solar Cycle 23) | | 1989 - 91 (Solar Cycle 22) | | 2000 - 02 (Solar Cycle 23) | | 1989 - 91 (Solar Cycle 22) | |
| | $\Delta I/\Delta P$ | $R$ | $\Delta I/\Delta P$ | $R$ | $\Delta I/\Delta P$ | $R$ | $\Delta I/\Delta P$ | $R$ |
| SSN | -2.01 ± 0.7 | -0.40 | -3.24 ± 0.97 | -0.47 | -1.39 ± 0.39 | -0.50 | -1.97 ± 0.54 | -0.51 |
| Tilt angle [degree] | -2.04 ± 1.83 | -0.18 | -7.4 ± 2.88 | -0.38 | -1.76 ± 1.0 | -0.27 | -4.30 ± 1.62 | -0.39 |
| $V$ [km s$^{-1}$] | -0.98 ± 0.47 | -0.31 | -1.78 ± 0.56 | -0.45 | -0.57 ± 0.26 | -0.33 | -0.91 ± 0.32 | -0.41 |
| $B$ [nT] | -68.03 ± 19.7 | -0.48 | -63.4 ± 20.2 | -0.45 | -33.84 ± 11.27 | -0.43 | -33.94 ± 11.52 | -0.43 |
| $\sigma_B$ [nT] | -55.9 ± 20.5 | -0.40 | -56.5 ± 21.8 | -0.38 | -27.87 ± 11.58 | -0.36 | -30.53 ± 12.36 | -0.37 |
| $BV/1000$ [mV m$^{-1}$] | -107.4 ± 30.3 | -0.49 | -106.6 ± 27.9 | -0.52 | -56.32 ± 17.10 | -0.47 | -56.14 ± 16.11 | -0.49 |
| $BV^2 \cdot 10^{-6}$ [$10^{-6}$ mV s$^{-1}$] | -154.4 ± 48.2 | -0.46 | -161.3 ± 41.4 | -0.53 | -82.1 ± 27.3 | -0.44 | -83.9 ± 23.7 | -0.49 |

i. CRI decreases at a faster rate with solar/interplanetary parameters in $A<0$ than $A>0$ during low, increasing and decreasing solar activity.
ii. During high solar activity (a period of mixed polarity), the rate of change in CRI with change in most of the solar/interplanetary parameters in almost same during two consecutive Solar Cycles 22 and 23.
iii. The CRI with solar-wind velocity, in particular, is correlated strongly in $A<0$ ($R = -0.80$) as compared to $A>0$ ($R = -0.61$) during low solar-activity epochs. During the decreasing activity phase, the correlation is comparatively much better in $A<0$ ($R = -0.87$) as compared in $A>0$ ($R = 0.12$) epochs.
iv. During low solar-activity conditions, the CRI response to changes in various parameters is two to three times or even more in $A<0$ as compared to $A>0$ polarity.
v. During decreasing and increasing solar-activity conditions the CRI response to changes in solar/interplanetary parameters is about 1.5 to 2 times more in $A<0$ as compared to $A>0$ polarity conditions.
vi. During high solar-activity conditions in Cycle 22 and 23, no significant difference in CRI-response to changes in various solar/interplanetary parameters is observed. Moreover, the correlations are also not good, in general and with tilt angle in particular, in this period.

The variable Sun controls the structure of the heliosphere and the modulation of cosmic rays through the level of solar activity, the tilt of the heliospheric current sheet, the velocity of the solar wind and the strength and turbulence of the interplanetary magnetic field (McDonald, Webber, and Reames, 2010). Determination of diffusion effects is a challenging astrophysical problem, because it requires an understanding of the properties of magnetic fields and turbulence, it also demands accurate theories for determining diffusion tensor (*e.g.* Pei *et al.*, 2012). The inverse correlation between CRI and interplanetary magnetic field [$B$] has been known for a long time (*e.g.* Burlaga and Ness, 1998; Cane *et al.*, 1999). Parallel diffusion coefficient [$K$] as well as the perpendicular diffusion coefficient [$K$] is assumed to be generally inversely proportional to $B$ in theoretical modeling of solar modulation (Jokipii and Davila, 1981; Reinecke, Moraal, and McDonald, 2000). Drift velocities of CRI increase with decreasing $B$. Thus, it would be interesting to look whether the $K$ and/or $K$ relation with $B$ shows any polarity-dependent effect, at least during low solar-activity periods, in view of our results showing the polarity-dependent CRI-response to changes in *B*.

Table 6 Ratio of $\Delta I/\Delta P$ in different polarity states, but similar solar activity conditions.

| Parameters | Low (minimum) $\Delta I/\Delta P$ | | Increasing $\Delta I/\Delta P$ | | Decreasing $\Delta I/\Delta P$ | | High (maximum) $\Delta I/\Delta P$ | |
|---|---|---|---|---|---|---|---|---|
| | Oulu NM (A<0):(A>0) (2007-09): (1995-97) | Newark NM (A<0):(A>0) (2007-09): (1995-97) | Oulu NM (A<0):(A>0) (2009-11): (1997-99) | Newark NM (A<0):(A>0) (2009-11): (1997-99) | Oulu NM (A<0):(A>0) (2003-05): (1992-94) | Newark NM (A<0):(A>0) (2003-05): (1992-94) | Oulu NM [2] (2000-02): (1989-91) | Newark NM [2] (2000-02): (1989-91) |
| SSN | 2.64 | 2.60 | 1.93 | 2.24 | 1.05 | 1.23 | 0.62 | 0.71 |
| Tilt angle [degree] | 2.92 | 3.21 | 1.2 | 1.38 | 1.58 | 1.41 | 0.28 | 0.41 |
| $V$ [km s$^{-1}$] | 2.03 | 1.90 | 1.03 | 1.19 | ---[1] | ---[1] | 0.55 | 0.63 |
| $B$ [nT] | 3.41 | 2.70 | 1.55 | 1.74 | 1.76 | 1.77 | 1.07 | 0.99 |
| $\sigma_B$ [nT] | 4.35 | 3.35 | 1.49 | 1.65 | 1.66 | 1.63 | 0.99 | 0.91 |
| $BV/1000$ [mV m$^{-1}$] | 5.05 | 2.40 | 1.45 | 1.78 | 1.34 | 1.42 | 1.01 | 1.00 |
| $BV^2 \cdot 10^{-6}$ [10$^{-6}$ mV s$^{-1}$] | 5.51 | 2.18 | 1.49 | 1.78 | 1.80 | 2.01 | 0.96 | 0.98 |

[1] Value of *R* is very poor during *A*>0 (<0.20)
[2] Mixed polarity periods

In the drift formulation of cosmic-ray modulation (Kota and Jokipii, 1983; Potgieter, 2013 and references therein) positively charged cosmic rays preferentially enter the heliosphere from the direction tied to the solar poles during *A*>0 periods (*e.g.* 1970 - 1980 and 1990 - 2000). During *A*<0 periods (*e.g.* 1960 - 1970 and 1980 - 1990), when the solar field polarity is reversed, cosmic rays approach the sun from the equatorial region along the HCS. In this case, cosmic rays are supposed to be more affected by magnetic structures characterised by enhanced magnetic fields (see *e.g.,* Laurenza *et al.,* 2012).

While drift effects are qualitatively well understood, the magnitude of drift effects remains, however, still uncertain and debated. The reason for this uncertainty is that, typically, the tilt of the HCS changes in phase with the solar activity and with other solar and heliospheric parameters such as the magnetic-field strength, hence the contribution of the different effects are difficult to separate (Kota, 2013). Analyses such as ours, involving various solar/heliospheric parameters in different polarity epochs and solar-activity conditions, are expected to improve our understanding of the unexplained aspects of solar modulation of galactic cosmic rays.

## 3. Conclusions

We find that the CRI decreases at a faster rate with an increase in both SSN and as well as with *V* and *B* and their derivatives [*BV*, *BV*$^2$] when the solar magnetic parameter *A* is negative than when *A* is positive. Thus not only SSN and but also the parameters *V* and *B* exhibit different effectiveness in modulating the CRI during *A*<0 as compared to *A*>0 solar-polarity epochs. This rate is found to be faster by a factor of 1.5 to 2 in *A*<0 than in *A*>0, during the increasing and decreasing phases of solar activity cycles and even two to three times faster in low activity periods. More specifically, we find that during *A*<0 increasing phase of solar cycles, CRI decreases at about twice the faster rates with an increase in SSN, ,

$B$, and $E$ while this rate is about 1.5 times higher with $V$, $_B$, and $BV^2$, as compared to rates during a similar phase of other solar cycles in $A>0$ epochs. There is essentially no consistent difference in CRI effectiveness to changes in various parameters during high solar activity in different solar cycles.

All four terms (convection, diffusion, drift, and adiabatic energy change) in Parker's transport equation involve $V$ and/or $B$ in one form or the other. Out of these four terms, only the curvature and gradient drifts are considered to be solar-polarity dependent. However, from our results it appears that the response of both $V$ and $B$ (which are included in convection/diffusion terms) to changes in CRI are solar-polarity dependent during all phases of solar cycle except solar maximum.

This observed difference in the CR-effectiveness of different parameters in $A<0$ and $A>0$ polarity epochs (in similar solar activity conditions) may be ascribed to different access routes of cosmic-ray particles in $A<0$ and $A>0$ polarity states as predicted by drift dominated models of cosmic ray modulation. However, the possibility of polarity-dependent effects on transport parameters needs to be explored.

**Acknowledgements** We thank Station Manager Ilya Usoskin and Sodankyla Geophysical Observatory for the online availability of Oulu-neutron monitor data. We also thank the National Science Foundation (supporting Bartol Research Institute neutron monitors) and Principle Investigator John W. Bieber for the online availability of Newark neutron monitor data. Availability of solar and plasma/field data through the NASA/GSFC OMNI Web interface and the HCS inclination data (courtesy of J.T. Hoeksema) are also acknowledged. We also thank the reviewer for useful and constructive comments.